\documentclass[aps,prl,reprint,groupedaddress,amsmath,amssymb,preprintnumbers,showpacs]{revtex4-1}

\usepackage{enumerate}
\usepackage{graphicx}
\usepackage{color}
\usepackage{float} 
\usepackage{xspace}
\graphicspath{{./}}

\newcommand{\mol}{$\text{H}_2^+$\xspace}
\newcommand{\floqel}[1]{\xi_{E_\text{e}}}
\newcommand{\floqnuc}[1]{\xi_{E_\text{N}}}

\newcommand{\en}{E_\text{N}}

\begin{document}
\title{Characterization of Molecular Breakup by very Intense, Femtosecond XUV Laser Pulses}
\author{Lun \surname{Yue}}
\author{Lars Bojer \surname{Madsen}}
\affiliation{Department of Physics and Astronomy, Aarhus University, DK-8000 Aarhus C, Denmark}
\date{\today}

\begin{abstract}
  We study the breakup of \mol exposed to super-intense, femtosecond laser pulses with frequencies greater than that corresponding to the ionization potential. By solving the time-dependent Schr\"{o}dinger equation in an extensive field parameter range, it is revealed that highly nonresonant dissociation channels can dominate over ionization. By considering field-dressed Born-Oppenheimer potential energy curves in the reference frame following a free electron in the field, we propose a simple physical model that characterizes this dissociation mechanism. The model is used to predict control of vibrational excitation, magnitude of the dissociation yields, and nuclear kinetic energy release spectra. Finally, the joint energy spectrum for the ionization process illustrates the energy sharing between the electron and the nuclei and the correlation between ionization and dissociation processes.

\end{abstract}

\pacs{33.80.Rv, 33.80.Gj, 82.50.Kx}

\maketitle

Molecular breakup processes induced by strong light-matter interactions are of fundamental interest. Molecular-specific ionization phenomena include charge-resonance-enhanced ionization \cite{Zuo95}, subcycle multiple ionization bursts \cite{Takemoto10}, electron-nuclei energy-sharing in above-threshold dissociative ionization \cite{Madsen12}, above-threshold Coulomb explosion \cite{Esry06}, and interplay between multiphoton and tunneling ionization \cite{Silva13}. Dissociation phenomena include above-threshold dissociation \cite{Giusti-Suzor90}, bond-softening \cite{Bucksbaum90}, bond-hardening \cite{Bandrauk81,Zavriyev93}, and rescattering induced dissociation \cite{Niikura02}. All these processes were discovered in the low-frequency regime where absorption of several photons is needed to reach a breakup channel. Depending on the laser parameters, either dissociation or ionization dominates \cite{Kastner09,Picon12,He12}.

With advancements in light-source technology,  extreme-ultraviolet (XUV) laser pulses of femto- and subfemtosecond duration are now produced from high-order harmonic \cite{Sansone11,Drescher01,Paul01,Hentschel01} or free-electron lasers \cite{Ackermann07,McNeil10}. New focusing techniques \cite{Mashiko04,Sorokin07,Nelson09} led to XUV femtosecond (fs) pulses with peak intensities $I\ge 4\times 10^{17}$ W/cm$^2$ \cite{Andreasson11}. Intense XUV pulses were, e.g., applied on rare gases to study sequential versus non-sequential multiple ionization \cite{Moshammer07,Rudenko08}, and creation of charge states up to 21 in Xe was observed \cite{Sorokin07}. For molecules, experiments on $\text{HeH}^+$ \cite{Pedersen07} and $\text{N}_2$ \cite{Jiang09} provided benchmark data for theory. The high-frequency regime is defined in this work as the regime where one-photon ionization is allowed. Since the photon resonance is much closer to the threshold for ionization than for dissociation, ionization is at first glance expected to dominate.

Here we characterize some hitherto unobserved molecular breakup phenomena in the regime of super-intense, high-frequency, fs pulses, supplementing the phenomena known from the low-frequency regime, and adding insight to the general field of strong laser-matter interaction.
The characteristics we find include
(a) even with full inclusion of nuclear dynamics, stabilization against ionization occurs, i.e., the ionization yield does not necessarily increase with intensity \cite{Gavrila02,Popov03};
(b) a mechanism by which dissociation, in contrast to the expectation from energy considerations, completely dominates over ionization;
(c) control over the vibrational distribution, dissociation yield, and nuclear kinetic energy release (NKER) spectra by the parameters of the laser pulse;
(d) insight into the energy sharing between electronic and nuclear degrees of freedom, as displayed by the joint energy spectrum (JES).

We consider $\text{H}_2^+$, a molecule that was used to discover many low-frequency processes that were later observed in more complex molecules (see, e.g., \cite{Madsen12,Wu13}).
The possible extension of our findings to other molecules is discussed in the end. We consider a co-linear model that includes the dimension that is aligned with the linearly polarized pulse. This model allows us to capture the essential physics, and to perform calculations over an extensive laser parameter regime.
The time-dependent Schr\"{o}dinger equation (TDSE) reads (atomic units are used throughout),
\begin{equation} 
  i\frac{\partial}{\partial t}\Psi(x,R,t)
  =\left[ T_\text{e}+T_\text{N}+V_\text{eN}(x,R)+V_\text{I}(t)\right]\Psi(x,R,t)
  \label{eq:tdse}
\end{equation}
with $x$ the electronic coordinate measured with respect to the center of mass of the nuclei, $R$ the internuclear distance, $T_\text{e}=-(1/2\mu)\partial^2/\partial x^2$, $T_\text{N}=-(1/m_\text{p})\partial^2/\partial R^2$, $V_\text{eN}(x,R)=-1/\sqrt{{(x-R/2)^2+a(R)}}-1/\sqrt{(x+R/2)^2+a(R)}+1/R$, and $V_\text{I}(t)=-i \beta A(t) \partial/\partial x $, with $m_\text{p}$ the proton mass, $\mu=2m_\text{p}/(2m_\text{p}+1)$, $\beta = (m_\text{p}+1)/m_\text{p}$, and $a(R)$ the softening parameter producing the exact $1s\sigma_g$ Born-Oppenheimer (BO) curve \cite{Madsen12}. The initial ground state has the energy $E_0=-0.5973$, the dissociation limit $E_\text{d}=-0.5$, the equilibrium internuclear distance $R_0=2.064$, and the ionization potential $I_\text{p}=1.1$ at $R_0$.
 We use vector potentials $A(t)=\left(F_0/\omega\right) g(t)\cos\omega t$, with $F_0$ related to the peak intensity by $I=F_0^2$. The envelope is $g(t)=\exp\left( -4\ln(2) t^2/\tau^2 \right)$, with $\tau$ the FWHM, and the number of cycles $N_\text{c}$ defined by $\tau=4\pi\sqrt{\ln2}N_\text{c}/\omega$. All pulses used satisfy the non-relativistic criteria $2U_p/c^2\ll 1$ \cite{Mourou06,Ludwig14} and the dipole condition $U_p/2\omega c \ll 1$ \cite{Reiss08}, with $U_\text{p}=F_0^2/4\omega^2$ the ponderomotive potential.

\begin{figure}
  \centering
  \includegraphics[width=0.45\textwidth]{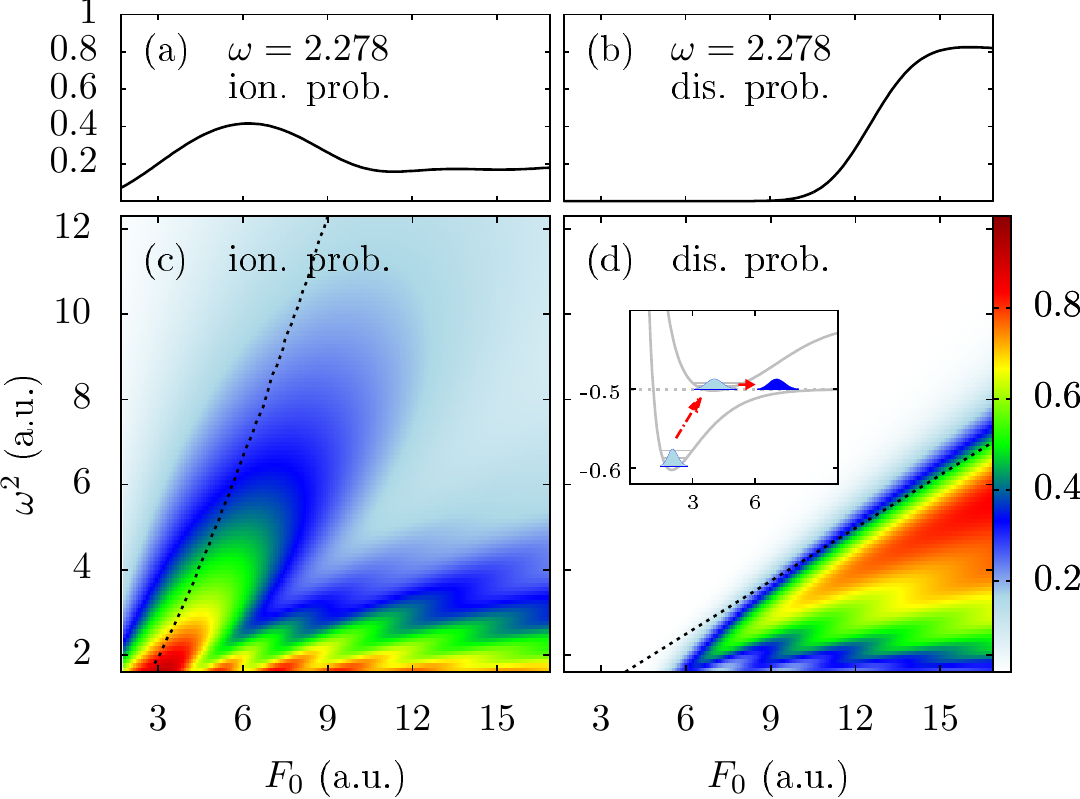}
  \caption{(color online). Continuum probabilities for \mol interacting with intense laser pulses containing 100 cycles, corresponding to $\tau=2.1 - 15.8$ fs. (a) and (b) DI and dissociation probabilities for $\omega=2.278$, (c) and (d) DI and dissociation probabilities for different $\omega^2$ and $F_0$. The dashed line in (c) traces the position of the largest DI rate, calculated from HFFT (see text). The dashed, straight line in (d) corresponds to $\alpha_0=F_0/\omega^2=2.41$. The inset in (d) shows the nuclear dynamics for pulse parameters tracing this line (see Fig.~\ref{fig2} and the accompanying discussion).}
  \label{fig1}
\end{figure}

We solve \eqref{eq:tdse} numerically \cite{Feit82} and obtain yields for the dissociation and dissociative ionization (DI) channels of H$_2^+$ (see Suppl. Material \cite{suppl14}).
Fig.~\ref{fig1}(a) shows the DI probability $P_\text{ion}$ of \mol 
resulting from laser pulses with $\tau=11.1$ fs and $\omega=2.278$. $P_\text{ion}$ increases with $F_0$ for lower amplitudes until it reaches a maximum at $F_0\simeq 6.2$, whereafter it decreases. The latter behavior indicates  stabilization with respect to DI \cite{Gavrila02}. Figure~\ref{fig1}(b) presents the dissociation probability $P_\text{dis}$. For the lower values of $F_0$, we observe no dissociation as expected from pertubation theory. At $F_0\simeq 9$, dissociation sets in, increasing with $F_0$, until it ``saturates'' at $F_0\simeq 15$.

To obtain a complete picture, Figs.~\ref{fig1}(c) and ~\ref{fig1}(d) present $P_\text{ion}$ and $P_\text{dis}$, as functions of $F_0$ and $\omega^2$. For fixed $\omega$ the suppression of DI for large $F_0$ is evident [Fig.~\ref{fig1}(c)]. Lobes corresponding to large $P_\text{ion}$ are seen emanating from the origin. By using a version of high-frequency Floquet theory (HFFT) \cite{Gavrila92} generalized to include nuclear motion \cite{suppl14}, we find that the largest lobe in Fig.~\ref{fig1}(c) is along the largest peak in the DI rates. The dashed line traces the largest HFFT rate, in reasonable agreement with the TDSE result. 
The oscillatory behavior in $P_\text{ion}$ was discussed for reduced-dimensionality atomic systems \cite{Gavrila02, Su97, Dorr00}, and was attributed to the two-center character of dressed the atomic wave function.
In the case of $P_\text{dis}$ in Fig.~\ref{fig1}(d), we observe appreciable dissociation for pulses with $\alpha_0 =F_0/\omega^2\geq 2.41$ indicated by the dashed line. In this regime, $P_\text{ion}+P_\text{dis}=1$, implying unity probability for breakup. The remnants of the lobes from DI are present in Fig.~\ref{fig1}(d), an indication of the different time scales for DI and dissociation processes: DI occurs during the pulse, and what is not ionized, dissociates after the pulse. The physics of the dashed line will be explained below.
 
\begin{figure}
  \centering
  \includegraphics[width=0.45\textwidth]{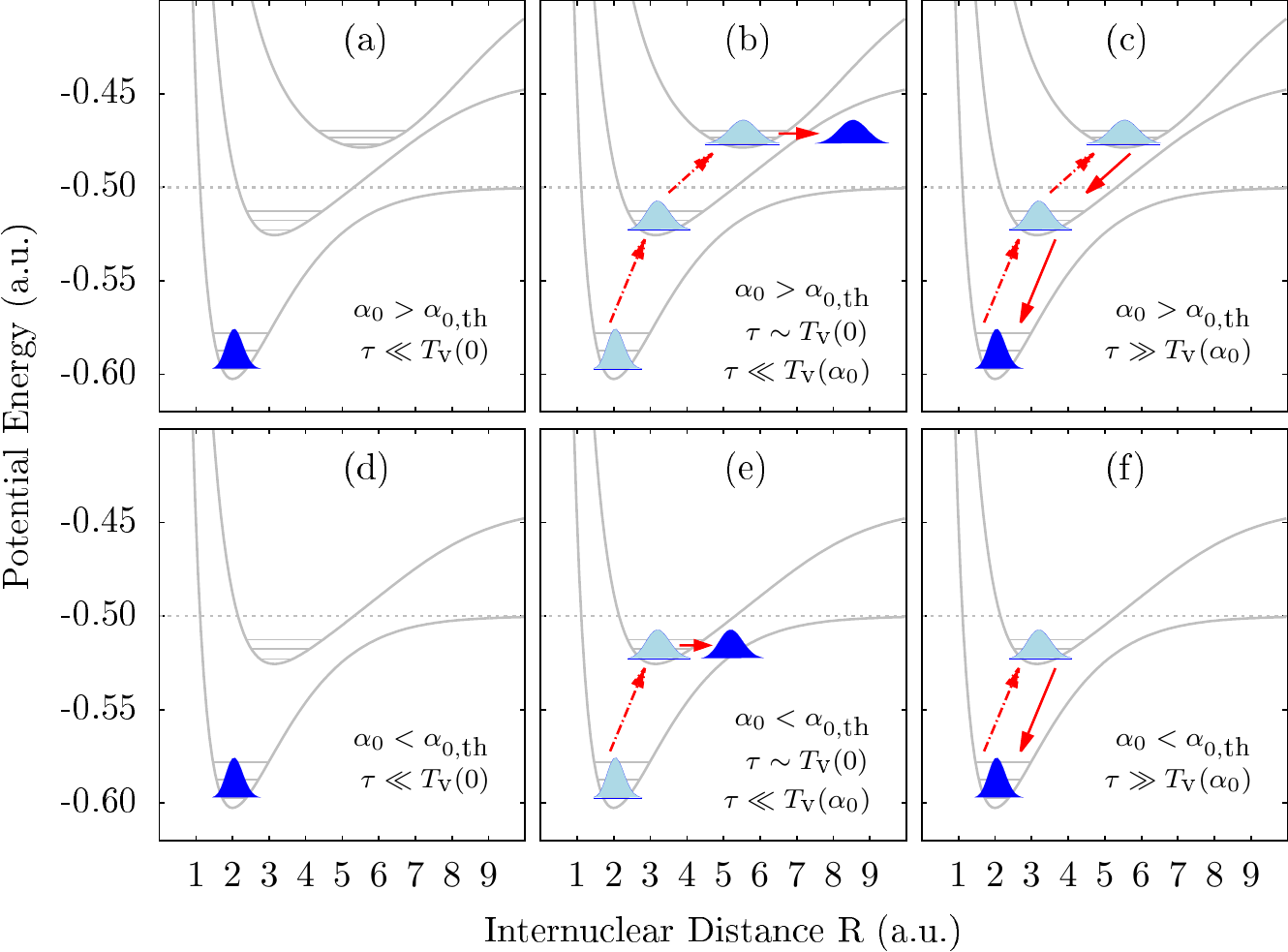}
  \caption{(color online). Schematic of the nuclear dynamics in the KH frame for different $\alpha_0$ and $\tau$. For (a)-(c), $\alpha_0=3.25$, and for (d)-(f), $\alpha_0=1.87$. In each panel, from bottom to top, the lowest field-dressed BO curve is shown for different times corresponding to $\alpha_0(t)=0$, $1.87$, (and $3.25$). The lowest three field-dressed vibrational levels are indicated. The horizontal lines indicate $E_\text{d}$. The arrows sketch the pathways of the vibrational WP during the turn-on and -off of the pulse.  The final position of the WP after the pulse is shown in dark blue.}
  \label{fig2}
\end{figure}

In order to elucidate the origin of the onset of dissociation in Fig.~\ref{fig1}, it is instructive to transform Eq.~\eqref{eq:tdse} into the Kramers-Henneberger (KH) frame \cite{Kramers56,Henneberger68}, where the complete laser-molecule interaction is contained in the modified electron-nuclei interaction $V_\text{eN}(x+\beta\alpha(t),R)$, with the quiver motion $\alpha(t)=\int^tA(t')dt'$. In the case of a monochromatic laser field, the Floquet ansatz for the wave packet (WP) and the Fourier expansion of $V_\text{eN}(x+\beta\alpha(t),R)$ results in a coupled set of equations. For large $\omega$, effectively only the zeroth order Fourier component, $V_0(x,R,\alpha_0)=(\omega/2\pi)\int_0^{2\pi/\omega}V_\text{eN}(x+\beta\alpha(t),R)dt$, remains, resulting in the structure equation \cite{Henneberger68,suppl14,Gavrila02}
\begin{equation}
  \begin{aligned}
    \left[H_\text{e}(x,R;\alpha_0)+T_\text{N}\right] u(x,R;\alpha_0)
    =W(\alpha_0)u(x,R;\alpha_0),
    \label{eq:structure}
  \end{aligned}
\end{equation}
with $H_\text{e}(x,R;\alpha_0)=T_\text{e}+V_0(x,R;\alpha_0)$ the field-dressed electronic Hamiltonian. For a given $\alpha_0$, Eq.~\eqref{eq:structure} is solved in the BO-approximation, yielding the field-dressed BO curves $E_\text{el,i}(R,\alpha_0)$ and the dressed energies $W_{i,\nu}(\alpha_0)$ \cite{Shertzer94}, with the indices $i=1,2,\dots$ and $\nu=0,1,...$ denoting the electronic and vibrational states. To treat pulsed laser fields, we let the maximal quiver amplitude vary with the field envelope, $\alpha_0\rightarrow\alpha_0(t)\equiv \alpha_0g(t)$. The lowest BO curve is plotted in Fig.~\ref{fig2} for $\alpha_0(t)=0,1.87$ and $3.25$. With increasing $\alpha_0(t)$, the BO curve is shifted upwards in energy, towards greater $R$, and becomes gradually shallower. The latter implies that the dressed vibrational time scale, $T_\text{v}(\alpha_0(t))\equiv 2\pi/\left[W_{1,1}(\alpha_0(t))-W_{1,0}(\alpha_0(t))\right]$, increases with $\alpha_0(t)$ (see also \cite{suppl14}).

We now present a qualitative model of the dissociation mechanism. The validity of the model is determined later by TDSE results. 
Let 
$\tau$ be
the time scale for the turn-on (and -off) of the laser pulse, and $\alpha_{0,\text{th}}$ the quiver amplitude satisfying $W_{1,0}(\alpha_{0,\text{th}})=E_\text{d}$, i.e., when the dressed ground state equals the dissociation limit [inset of Fig.~\ref{fig1}(d)]. 
Provided the pulse satisfies 
(i)~$\alpha_0 > \alpha_{0,\text{th}}$,
(ii)~$T_\text{v}(0)\sim \tau$, and
(iii)~$\tau \ll T_\text{v}(\alpha_0)$, the dissociation process occurs
as follows [Fig.~\ref{fig2}(b)]. During the turn-on of the pulse, (ii) ensures the population to follow the field-dressed ground state adiabatically. At the field maximum, (i) implies that the bound WP populates dressed eigenstates with energies greater than $E_\text{d}$. Due to (iii), the turn-off of the pulse can be considered sudden, and the nuclear WP does not feel the fast change of the electronic potential, leaving its position and energy unchanged. After the pulse, the nuclear WP is trapped above $E_\text{d}$, resulting in dissociation via the field-free electronic ground state, with NKER given by $E_\text{N}(\alpha_0)=W_{1,0}(\alpha_0)-E_\text{d}$.

The laser field regime for which $P_\text{dis}$ is nonzero in Fig.~\ref{fig1}(d) satisfies (i)-(iii).
For \mol, we have $\alpha_{0,\text{th}}=2.41$, $T_\text{v}(0)=15.2$ fs and $T_\text{v}(\alpha_{0,\text{th}})=41.7$ fs. In Fig.~\ref{fig1}(d), the onset of dissociation is  indeed around the dashed line corresponding to $\alpha_0=\alpha_{0,\text{th}}$. This agreement supports the physical picture of the model. The frequency range $\omega^2=2-8$ in Fig.~\ref{fig1}(d) where $P_\text{dis}$ is nonzero corresponds to $\tau=9.0-17.9$ fs, which fulfills condition (ii) at least approximately. For $\alpha_0 > \alpha_{0,\text{th}}$, we have $\tau\ll T_\text{v}(\alpha_{0,\text{th}}) < T_\text{v}(\alpha_0)$, and condition (iii) is satisfied as well. DI occurs throughout the whole duration of the pulse due to higher-order Floquet components in $V_\text{eN}(x+\beta\alpha(t),R)$. In the stabilization regime, DI is greatly suppressed, leaving the population trapped above $E_\text{d}$ to dissociate. This explains $P_\text{ion}+P_\text{dis}=1$ in the parameter regime of Fig.~\ref{fig1}(d) where $P_\text{dis}$ is nonzero. 

Figures~\ref{fig2}(a) and (c) illustrate the cases for which (i) is satisfied, but (ii) and (iii) are not. In the case where the pulse duration is short, $\tau\ll T_\text{v}(0)$, the perturbation can be considered sudden, and the initial population is unaffected, resulting in no dissociation [Fig.~\ref{fig2}(a)]. For long pulses $\tau \gg T_\text{v}(\alpha_0)$, the adiabatic approximation is accurate, and the initial state follows the dressed ground state throughout the whole pulse resulting in a final bound distribution similar to the initial population [Fig.~\ref{fig2}(c)]. For the case where (ii) and (iii) are satisfied, but with $\alpha_0<\alpha_{0,\text{th}}$, no dissociation occurs. Instead, an excited vibrational WP is created containing field-free vibrational states with quantum numbers $\nu$ satisfying $W_{1,\nu}(0)=W_{1,0}(\alpha_0)$ [Fig.~\ref{fig2}(e)].

\begin{figure}
  \centering
  \includegraphics[width=0.45\textwidth]{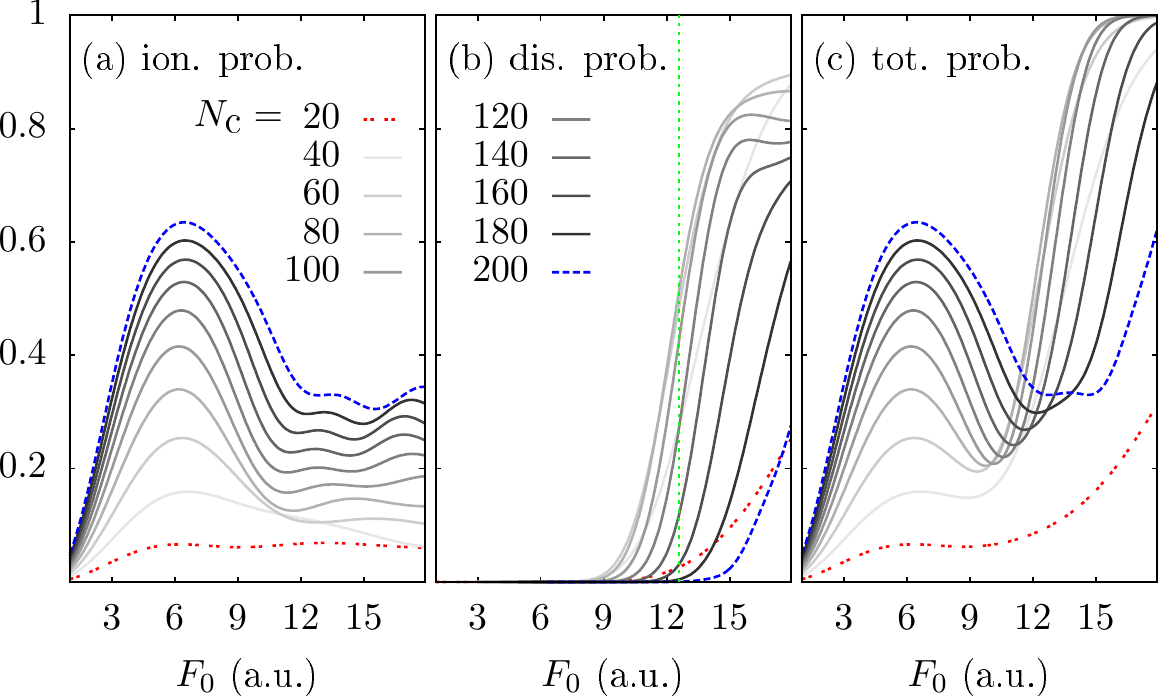}
  \caption{(color online).
    Vibrational distributions at the end of the pulse (bottom panels), and NKER spectra for dissociation in the ground electronic $1s\sigma_g$ state of H$_2^+$ (top panels). The laser pulses have $\omega=2.278$, varying $F_0$, and different $N_c$.
 The vertical axes are equidistant in energy using the same scale for the bound (bottom panels) and continuum (top panels) parts. Each row has the same color scale. The dotted lines indicate the energies of the field-dressed vibrational ground state at field maxima with respect to the dissociation limit, $W_{1,0}(\alpha_0)-E_\text{d}$.
}
  \label{fig3}
\end{figure}

The mechanisms described in Fig.~\ref{fig2} imply that by varying $\tau$ and $\alpha_0$, the vibrational populations, dissociation yields, and NKER can be controlled. This prediction is confirmed by the TDSE results of Fig.~\ref{fig3}, which shows the bound vibrational populations and the NKER spectra for the $1s\sigma_g$ state at the end of the pulse, for $\omega=2.278$, $F_0=1-17.9$, and $N_c=10-500$. We checked that no dissociation occurs via the excited electronic states.
For $N_\text{c}=10$, the $\nu=0$ level is most populated, except at $F_0\gtrsim 14$ where strong non-adiabatic couplings excite some vibrations. No dissociation occurs, in agreement with Fig.~\ref{fig2}(a). For $N_\text{c}=140$, the vibrational population and the NKER spectra follow approximately the dotted line corresponding to $W_{1,0}(\alpha_0)-E_\text{d}$. This is in agreement with Figs.~\ref{fig2}(b), \ref{fig2}(e), and the accompanying discussions, where we argued that the energy of the final WP with respect to $E_\text{d}$ after the pulse is $W_{1,0}(\alpha_0)-E_\text{d}$. If $W_{1,0}(\alpha_0)<E_\text{d}$, the WP is bound, and field-free vibrational states satisfying $W_{1,\nu}(0)\simeq W_{1,0}(\alpha_0)$ are populated. If $W_{1,0}(\alpha_0)>E_\text{d}$, the WP dissociates with NKER $E_\text{N}\simeq W_{1,0}(\alpha_0)-E_\text{d}$. 
Note that the onset of dissociation occurs at $F_0 \simeq \omega^2\alpha_{0,\text{th}}$. From $N_\text{c}=140$ to $N_\text{c}=500$, the adiabatic approximation for the evolution of the WP described in Fig.~\ref{fig2}(c) becomes gradually more appropriate, with the field-free vibrational ground state populated up to increasingly larger $F_0$. For $N_\text{c}=350-500$, no dissociation occurs. By suitable choices of $\tau$ and $F_0$, we can thus control the final vibrational populations and the NKER spectra. 
 
\begin{figure} 
  \centering
  \includegraphics[width=0.45\textwidth]{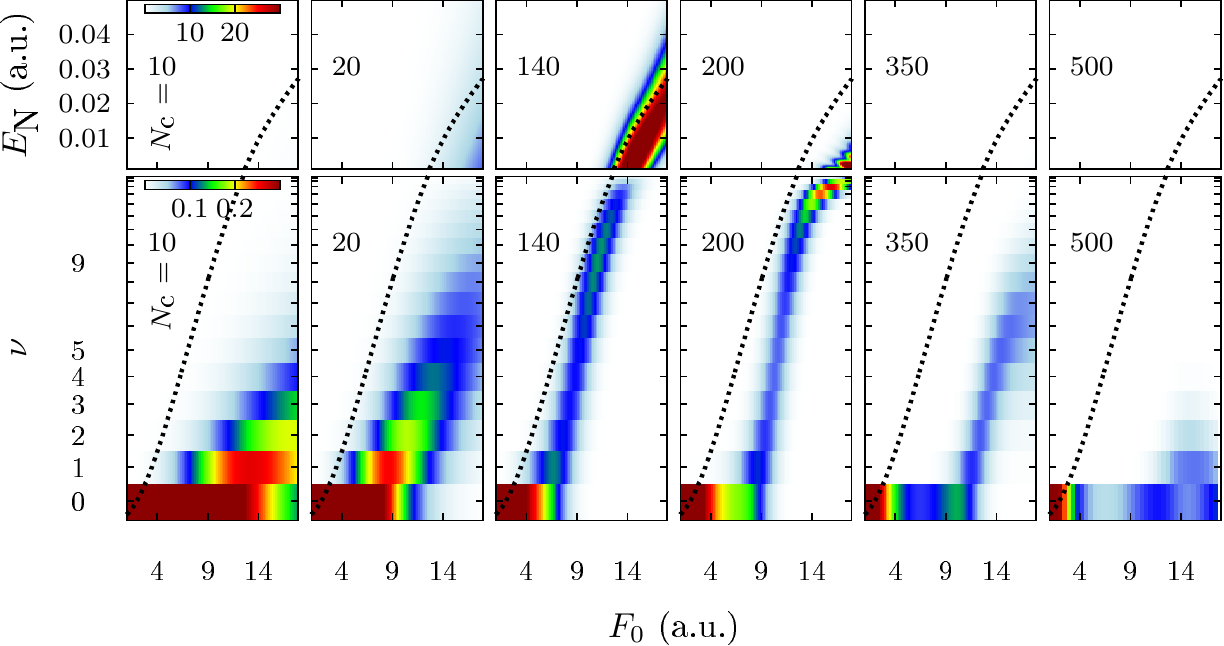}
  \caption{(color online). Time-resolved analysis showing the buildup of the JES for DI during a laser pulse
with parameters $\alpha_0=2.41$, $\omega=2.278$, and $N_\text{c}=100$. Diagonal lines indicate the energy conservations $E_\text{e}+E_\text{N}=E_0+\omega$ (left lines) and $E_\text{e}+E_\text{N}=W_{1,0}(\alpha_0)+\omega$. The upper and side subpanels show respectively the photoelectron and NKER spectra. $\text{Q}_1,\dots,\text{Q}_4$ denote the four maxima.}
  \label{fig4}
\end{figure}

To understand the interplay between DI and dissociation in this dynamical regime more deeply, Fig.~\ref{fig4} presents the formation of the JES describing the probability of measuring electronic $E_\text{e}$ and nuclear $E_\text{N}$ energies in the DI process, determined as in \cite{Yue13}.
Note that compared to Fig.~\ref{fig3}, the NKER energies from DI are much greater than that from dissociation.
The JES after the pulse ($t=\infty$) shows four distinct peaks, denoted by $\text{Q}_1,\dots,\text{Q}_4$. 
The physical picture in Fig.~\ref{fig2}(b) explains these.
At $t=-6.7$ fs, the Stark shift $\Delta(t) \equiv W_{1,0}(\alpha_0(t))-E_0$ is negligible, and ionization produces $\text{Q}_1$ in the JES along the line corresponding to the one-photon resonance $E_\text{e}+E_\text{N}=E_0+\omega$. At $t=0$, according to the model, the nuclear WP has the largest Stark shift $\Delta(0)=0.09732$, and ionization leads to $\text{Q}_2$ along the shifted one-photon resonance $E_\text{e}+E_\text{N}=W_{1,0}(\alpha_0)+\omega$. At $t>0$, the population stays above $E_\text{d}$ and dissociates. During dissociation, ionization occurs at large $R$, leading to $\text{Q}_3$ and $\text{Q}_4$ with low $\en$ in the JES at $t=6.7$. The structure of Q$_3$ and Q$_4$ will not be discussed here. Note that $\text{Q}_1$ is unchanging in magnitude and position for $t>0$, consistent with the model prediction that no field-free bound states become repopulated during pulse turn-off. For this reason the dynamic interference effect \cite{Demekhin12,Yue14}, where ionized WPs with equal continuum energy created during the rising and falling edges of the pulse interfere, is not observed.
These insights indicate that dissociation and probing of dissociation by DI can be achieved using a single pulse, with the nuclear WP being promoted to the field-free dissociation continuum during turn-on (pump) and probed through DI during turn-off.

The dynamics in terms of the physical model hinges on two requirements: a) stabilization with respect to DI; and b) increase of the $E_{\text{el},1}$ and $T_\text{v}$ with $\alpha_0$. Consider first a). We expect stabilization to occur in two-electron molecules such as H$_2$, D$_2$, and HeH$^+$, provided $\omega>I_\text{p}$ \cite{Gavrila96, Birkeland10}. For other diatomics like N$_2$, O$_2$, CO, or NO, the orbital energies can be grouped into two classes formed by innershells with energies less than -11, and outershells with energies larger than -2. The same classification holds for triatomic molecules (e.g., CO$_2$ and H$_2$O), and even larger systems like naphthalene.
For typical $\omega$'s considered in this work, e.g. 2-3, the outershells satisfy $\omega>I_{p}$ where we expect stabilization, while innershell electrons can be excited to Rydberg states by multiphoton absorption and subsequently be stabilized \cite{Gavrila02,Popov03}.
Now consider b). For large $\alpha_0$, it has been shown that $E_{\text{el},1}\propto \alpha_0^{-2/3}$ \cite{Gavrila96,Nguyen00}, while a perturbative analysis for small $\alpha_0$ \cite{suppl14}
shows that for a general diatomic molecule, $E_\text{el,1}$, $R_0$, and $T_\text{v}$ increases with $\alpha_0$. For intermediate $\alpha_0$, using quantum chemistry codes, the requirement b) was also fulfilled for H$_2$ \cite{Valle98,Valle97,Nguyen00}.
Even when the molecule is not aligned with the linearly polarized field, the mentioned two requirements a) and b) still holds \cite{Shertzer94,Nguyen00}, indicating that the observed dynamics would also occur in non-aligned molecules.
In light of these considerations, the described dynamics are expected to be observable in two-electron diatomics, while further studies are required to determine whether this is generally the case.
For a molecule X$_2$, with X some atomic species, the charged fragments X$^+$ and $e$ from DI can be detected in coincidence by cold-target recoil ion momentum spectroscopy \cite{Ullrich03}.
 The resulting JES will contain structures satisfying $E_\text{e}+2E_{\text{X}^+}=W_{1,0}(\alpha_0)+\omega$ for outershell DI, and  $E_\text{e}+2E_{\text{X}^+}=W_{1,0}(\alpha_0)+n_\text{min}\omega$ for innershell DI, with $n_\text{min}$ the minimum number of photons required. 
A large spread in $E_{\text{X}^+}$ containing low-energy X$^+$ fragments is a signature of continuous ionization during dissociation towards larger $R$, and thus a direct indication of the dressed dissociation channel (see Q$_2$-Q$_4$ in Fig.~\ref{fig4} for H$_2^+$).
We believe that an experimental verification would be challenging, but possible in the future, involving the generation of relevant pulses, molecular alignment, and channel-specific measurement techniques.

This work was supported by the Danish Center for Scientific Computing, an ERC-StG (Project No. 277767 - TDMET), and the VKR center of excellence, QUSCOPE.

\end{document}